\documentclass[12pt]{book}

\usepackage[dvips]{graphicx,color}
\usepackage{makeidx,tsukuba}

\makeauthorindex
\makeindex

\newcommand{\lsim}{\mathrel{\mathop{\kern 0pt \rlap
{\raise.2ex\hbox{$<$}}}
\lower.9ex\hbox{\kern-.190em $\sim$}}}
\newcommand{\gsim}{\mathrel{\mathop{\kern 0pt \rlap
{\raise.2ex\hbox{$>$}}}
\lower.9ex\hbox{\kern-.190em $\sim$}}}

\begin{document}

\BookTitle{\itshape The 28th International Cosmic Ray Conference}
\CopyRight{\copyright 2003 by Universal Academy Press, Inc.}
\pagenumbering{arabic}

\chapter{
Cosmic Ray Antiprotons from Relic Neutralinos in a Diffusion Model}

\author{%
%
%
Fiorenza Donato$^1$, Nicolao Fornengo$^1$, David Maurin$^2$, 
Richard Taillet$^3$ and Pierre Salati$^3$ \\
{\it (1) Dipartimento di Fisica Teorica,, University of Torino, 
10125 Torino, Italy\\
(2) Service d'Astrophysique, SAp CEA-Saclay, 91191 Gif-sur-Yvette
cedex, France\\
(3) LAPTH and Universit\'e de Savoie, 74941 Annecy-le-Vieux, France \\}
}

\section*{Abstract}
We use the constraints on the diffusion parameters as obtained with stable
nuclei to calculate the cosmic antiproton flux from annihilating  relic
neutralinos. We discuss the relevance of each characteristic parameter,
describing our two dimension diffusion model, on the flux of antiprotons
produced in the dark halo of our Galaxy.
We estimate a two orders of magnitude uncertainty on the
flux due to the unknowledge of the propagation parameters.
A conservative and systematic evaluation of the flux in the supersymmetric
parameter space
is done in order to exclude configurations providing a total (secondary
plus primary) flux in excess with observations.
We also study the effect on the flux induced by modifications in the
distribution of cold dark matter in the Galaxy.

\section{Introduction}
One of the most promising candidate for solving - at least partially - 
the problem of the astronomical dark matter comes from particle physics. 
In an extension of the Standard Model for elementary particles to 
supersymmetry, one can find a neutral and stable particle 
- the neutralino - perfectly fit to be a relic from the big bang. 
As well as providing good values for its relic density, the neutralino 
may be detected on Earth both directly and indirectly. 
Indirect detection lies on the measurement of its annihilation products as 
exotic components in rare cosmic fluxes, such gamma, neutrino, positron, 
antiproton, or antideuteron fluxes. 
In these Proceedings we present new results about the propagation of 
primary antiprotons - deriving from neutralino annihilation - in our 
Galaxy, by using previous studies on stable nuclei fluxes. 

\section{The neutralino induced antiproton flux}

\begin{figure}[h]
\vspace{-1.5cm}
  \begin{center}
    \includegraphics[height=23.5pc]{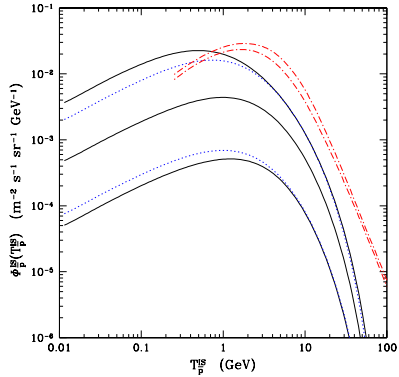}
  \end{center}
  \vspace{-3.5pc}
  \caption{The solid lines represent the antiproton flux for a 
$m_\chi$=100 GeV neutralino and for  maximal, 
median and minimal astrophysical configurations, for
$\chi^2_{B/C}\leq 40$. Dotted lines: the same, but for for 
$\chi^2_{B/C}\leq 30$. The dot--dashed band corresponds
to the secondary flux as taken from  [3]
for all the configurations giving $\chi^2_{B/C}\leq 40$.}
\label{fig:various_chi2}
\end{figure}

\noindent
The supersymmetric antiproton production differential rate,  
per unit volume and time is defined as
\begin{equation}
q_{\bar p}^{\rm susy}(r,z,T_{\bar p}) =
\, <\sigma_{\rm ann} v> \, g(T_{\bar p})
\left( \frac{ \rho_\chi (r,z)}{m_\chi} \right)^2 ,
\label{eq:source}
\end{equation}
where $<\sigma_{\rm ann} v>$ denotes the average over the galactic velocity
distribution function of the neutralino pair annihilation cross section
$\sigma_{\rm ann}$ multiplied by its relative velocity $v$, 
and $m_\chi$ is the  neutralino mass.  $g(T_{\bar p})$ 
denotes the $\bar p$ differential spectrum deriving from the hadronization
of quarks and gluons and  $\rho_\chi (r,z)$ is the mass distribution
function of neutralinos in the galactic halo.
All the quantities depending on the supersymmetric parameters have been
calculated in the framework of the eff{\sc mssm} [1].
 In order to obtain the distributions
$dN^h_{\bar p}/dT_{\bar p}$ the hadronization of quarks and gluons has been
evaluated by using the Monte Carlo code Jetset 7.2 [2].
\\
The propagation of cosmic rays in the Galaxy has been considered
in the framework of a two-zone diffusion model, which has been
described at length in [1,3]. 
The sources of primary antiprotons are distributed throughout the 
whole diffusive halo, and the solution to the transport equation
at our location (z=0) is given by [4]
\begin{equation}
     N_i(0) =   \int_{-L}^L \frac{q^{\bar{p},prim}_i(z)}{A_i}
     \frac{\displaystyle  \sinh \left(\frac{S_i(L-|z|)}{2}\right) 
}{\displaystyle \sinh \left(\frac{S_iL}{2}
     \right)}
      \, e^{ -V_c |z|/2K} dz
      \label{sol_prim}
\end{equation}
and where the energy-dependent quantities $S_{i}$ and $A_i$ are defined as
\begin{displaymath}
     S_i \equiv \left\{ \frac{V_c^2}{K^2}  +
     4  \frac{\zeta_i^2}{R^2} \right\}^{1/2} \hspace{1cm} \,\,\,
     A_i \equiv 2 \, h \, \Gamma^{ine}_{\bar{p}}
     \; + \; V_{c} \; + \; K \, S_{i} \,
     {\rm coth} \left\{ {\displaystyle \frac{S_{i} L}{2}} \right\}.
\end{displaymath}
The $q^{\bar{p},prim}_i(z, E)$ are the Bessel transforms of the source term
given by Eq. \ref{eq:source}.
\\
In Fig. \ref{fig:various_chi2} we present the results for the primary 
antiproton flux for $m_\chi$=100 GeV and $<\sigma v> = 2.3 \cdot 10^{-9}$ 
GeV$^{-2}$.
The band represented by solid lines corresponds to the maximal and 
minimal flux obtained from all the astrophysical configurations 
compatible with the analysis on stable nuclei and providing
$\chi^2_{B/C}\leq 40$ [5]. The two dotted lines give the maximal and 
minimal flux for the configurations with  $\chi^2_{B/C}\leq 30$. 
 We also plot the secondary 
antiproton flux as taken from [3 when all the 
configurations giving $\chi^2_{B/C}\leq 40$ are considered. 
We note the huge uncertainty in the 
calculation of the primary flux due to astrophysical parameters. 
The conservative choice of $\chi^2_{\rm B/C} < 40$ reflects in an 
uncertainty band two orders of magnitude large for energies 
 $T_{\bar{p}}\lsim 1$ GeV.
The two curves obtained  with parameters compatible with 
$\chi^2_{\rm B/C} < 30$ 
differs by a factor around 30 over all the energy range, just 
as the band  for $\chi^2_{\rm B/C} < 40$ at energies greater than 
 $T_{\bar{p}}\sim 1$ GeV.

\section{Comparison with data}
\begin{figure}[h]
\vspace{-2.5cm}
  \begin{center}
  \includegraphics[height=23.5pc]{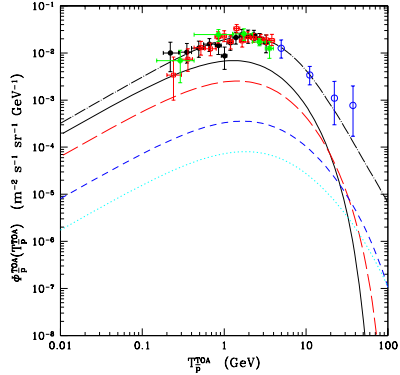}
  \end{center}
  \vspace{-5.5pc}
\caption{Primary flux for $m_\chi$= 60, 100, 300, 500 GeV (from top to bottom) 
obtained for the median astrophysical configuration.
The upper dot--dashed curve corresponds to the 
secondary flux taken from [3].
Full circles [6], open squared [7];
empty circles [8], starred [9].
\label{fig:prim_sec_data_min_solmin}}
 \end{figure}

Data on antiproton at Earth are now abundant, mostly because of the 
missions of the balloon borne detector {\sc bess}.
In Fig. \ref{fig:prim_sec_data_min_solmin}
 we compare our theoretical evaluations with data taken at 
solar minimum by {\sc bess} [6, 7], {\sc caprice} [8], and {\sc ams} [9].
The propagation parameters are the ones giving the best $\chi^2_{\rm B/C}$ and 
may be considered as the average ones, with respect to the incertitude band 
(see previous figure). 
We have calculated the flux for four different values of the neutralino mass:
 $m_\chi$= 60, 100, 300, 500 GeV.
We refer to [1] for all the details about both the astrophysical 
and supersymmetric aspects of the calculation. 
The lower mass gives the higher flux, even if for the most massive neutralinos 
the high energy part of the spectrum rises. 
We performed a full scanning of the supersymmetric parameter space 
[1]: using the median astrophysical propagation parameters no supersymmetric 
configuration may be excluded with present data and secondary flux extimation.

\section{Conclusions}
We have calculated the antiproton flux deriving from relic neutralino 
annihilation in the dark halo of our Galaxy. The diffusion model is the 
one previously selected by analysis on stable nuclei. 
Uncertainties due to propagation seriously affect the theoretical flux.
Comparison with data and secondary antiproton flux demonstrates that 
it is rather difficult to put constraints on the supersymmetric parameter
space. 
However, our results have been obtained assuming a spherical 
isothermal distribution of dark matter in the halo. Different 
density profiles or hypothesis about clumpy haloes have been proposed 
in the literature. They would lead 
to differencies in the primary flux  which may depend on the specific 
propagation model [1].

\section{References}

\vspace{\baselineskip}
\re
1.\ Donato F. et al. 2003, to appear (preprint DFTT-32-02)
\re
2.\ Tj\"{o}strand T.1994, Comput. Ph. Commun. 82, 74
\re
3.\ Donato F. et al. 2001, ApJ 563, 172 
\re
4.\ Barrau A. et al. 2002, A\&A 388, 676
\re
5.\  Maurin D., Donato F., Taillet R., Salati P. 2001, ApJ 555, 585
\re
6.\ Orito  S. et al. ({\sc bess} Collaboration) 2000, Phys. Rev. Lett. 84, 1078
\re
7.\ Maeno T. et al. ({\sc bess} Collaboration) 2001, Astropart. Phys. 16, 121
\re
8.\ Boezio M. et al. ({\sc caprice} Collaboration) 2001,  ApJ 561, 787
\re
9.\ Aguilar M. et al. ({\sc ams} Collaboration) 2002,  Phys. Rep. 366, 331

\endofpaper
\end{document}